\documentstyle[aps,twocolumn]{revtex}

\begin{document}

\title{Quantum state transfer and entanglement distribution among distant
nodes in a quantum network}
\author{J. I. Cirac$^{1,2}$, P. Zoller$^{1,2}$, H. J. Kimble$^{1,3}$, and H.
Mabuchi$^{1,3}$}
\address{$^1$ Institute for Theoretical Physics,
University of California at Santa Barbara, Santa Barbara, CA 93106-4030}
\address{$^2$ Institut f{\"u}r Theoretische Physik\\
Universit{\"a}t Innsbruck, Technikerstra{\ss}e 25, A-6020 Innsbruck, Austria}
\address{$^3$ Norman Bridge Laboratory of Physics 12-33, California Institute
of Technology, Pasadena CA 91125}
\date{\today}
\maketitle

\begin{abstract}
We propose a scheme to utilize photons for ideal quantum transmission
between atoms located at {\em spatially-separated} nodes of a quantum
network.  The transmission protocol employs special laser pulses which
excite an atom inside an optical cavity at the sending node so that its
state is mapped into a {\em time-symmetric} photon wavepacket that will
enter a cavity at the receiving node and be absorbed by an atom there
{\em with unit probability}.  Implementation of our scheme would enable
reliable transfer or sharing of entanglement among spatially distant atoms.
\end{abstract}


We consider a quantum network consisting of spatially separated nodes
connected by quantum communication channels.  Each node is a quantum
system that stores quantum information in quantum bits and processes
this information locally using quantum gates \cite{DiV95}. Exchange of
information between the nodes of the network is accomplished via
quantum channels.  A physical implementation of such a network could
consist {\it e.g.} of clusters of trapped atoms or ions representing
the nodes, with optical fibers or similar photon ``conduits''
providing the quantum channels.  Atoms and ions are particularly well
suited for storing qubits in long-lived internal states, and recently
proposed schemes for performing quantum gates between trapped atoms or
ions provides an attractive method for local processing within an
atom/ion node \cite{Ci95,Pe95,Mo95}.  On the other hand, photons
clearly represent the best qubit-carrier for fast and reliable
communication over long distances \cite{Tu95,Ma96}, since fast and
internal-state-preserving transportation of atoms or ions seems to be
technically intractable.

To date, no process has actually been identified for using photons (or
any other means) to achieve efficient {\em quantum transmission} between
spatially distant atoms.  In this letter we outline a scheme to implement
this basic building block of communication in a distributed quantum network.
Our scheme allows quantum transmission with (in principle) unit efficiency
between distant atoms 1 and 2 (see Fig.~1).  The possibility of
combing local quantum processing with quantum transmission between the
nodes of the network opens the possibility for a variety of novel
applications ranging from entangled-state cryptography\cite{Be95},
teleportation\cite{Be93} and purification \cite{Be96}, and is
interesting from the perspective of distributed quantum
computation\cite{magic}.

The basic idea of our scheme is to utilize strong coupling between a
high--Q optical cavity and the atoms\cite{Tu95} forming a given node
of the quantum network.  By applying laser beams, one first transfers
the internal state of an atom at the first node to the optical state
of the cavity mode.  The generated photons leak out of the cavity,
propagate as a wavepacket along the transmission line, and enter an
optical cavity at the second node.  Finally, the optical state of the
second cavity is transferred to the internal state of an atom.
Multiple-qubit transmissions can be achieved by sequentially
addressing pairs of atoms (one at each node), as entanglements between
arbitrarily located atoms are preserved by the state-mapping process.

The distinguishing feature of our protocol is that by
controlling the atom-cavity interaction, one can absolutely avoid the
reflection of the wavepackets from the second cavity, effectively
switching off the dominant loss channel that would be responsible for
decoherence in the communication process.  For a physical picture of
how this can be accomplished, let us consider that a photon leaks out
of an optical cavity and propagates away as a wavepacket.  Imagine
that we were able to ``time reverse'' this wavepacket and send it back
into the cavity; then this would restore the original (unknown)
superposition state of the atom, provided we would also reverse the
timing of the laser pulses. If, on the other hand, we are able to
drive the atom in a transmitting cavity in such a way that the
outgoing pulse were already symmetric in time, the wavepacket entering
a receiving cavity would ``mimic'' this time reversed process, thus
``restoring'' the state of the first atom in the second one.

The simplest possible configuration of quantum transmission between
two nodes consists of two two-level atoms 1 and 2 which are strongly
coupled to their respective cavity modes (see Fig.~1). The Hamiltonian
describing the interaction of each atom with the corresponding cavity
mode is ($\hbar=1$):
\begin{eqnarray}
\hat H_{i} &=& \omega_c \hat a_i^\dagger \hat a_i
+ \omega_0 |r\rangle_i\,{}_i\langle r|  +
g (|r\rangle_i\,{}_i\langle g| \hat a_i + h.c.)\nonumber\\
&& +  \frac{1}{2} \Omega_i(t) \left[ e^{-i[\omega_L t+\phi_i(t)]}
|r\rangle_i\,{}_i\langle e| + h.c \right] \quad (i=1,2).
\end{eqnarray}
Here, $\hat a_i$ is the destruction operator for cavity mode $i$ with
frequency $\omega_c$, $|g\rangle,|r\rangle$, and $|e\rangle$ form a
three--level system of excitation frequency $\omega_0$ (Fig.~1), and
the qubit is stored in a superposition of the two degenerate ground
states. The states $|e\rangle$ and $|g\rangle$ are coupled by a Raman
transition \cite{Pe95,Mo95,La96}, where a laser of frequency
$\omega_L$ excites the atom from $|e\rangle$ to $|r\rangle$ with a
time dependent Rabi frequency $\Omega_i(t)$ and phase $\phi_i(t)$,
followed by a transition $|r\rangle \rightarrow |e\rangle$ which is
accompanied by emission of a photon into the corresponding cavity
mode, with coupling constant $g$. In order to suppress spontaneous
emission from the excited state during the Raman process, we assume
that the laser is strongly detuned from the atomic transition
$|\Delta|\gg \Omega_{1,2}(t), g, |\dot \phi_{1,2} |$ (with
$\Delta=\omega_L-\omega_0$). In such a case, one can eliminate
adiabatically the excited states $|r\rangle_i$. The new Hamiltonian
for the dynamics of the two ground states becomes, in a rotating frame
for the cavity modes at the laser frequency,
\begin{eqnarray}
\hat H_{i} &=& - \delta \hat a_i^\dagger \hat a_i + \frac{g^2}{\Delta}
\hat a_i^\dagger \hat a_i |g\rangle_i\;_{i}\langle g| +
\delta\omega_i(t) |e\rangle_i\;_{i}\langle e| \nonumber\\ && - i
g_i(t)\left[e^{i\phi_i(t)} |e\rangle_i\;{}_i\langle g| a_i - {\rm
h.c.}  \right].  \quad (i=1,2)
\end{eqnarray}
The first term involves the Raman detuning $\delta=\omega_L-\omega_c$.
The next two terms are AC--Stark shifts of the ground states
$|g\rangle$ and $|e\rangle$ due to the cavity mode and laser field,
respectively, with $\delta\omega_i(t)=\Omega_i(t)^2/(4\Delta)$. The
last term is the familiar Jaynes--Cummings interaction, with an
effective coupling constant $g_i(t)= g\Omega_i(t)/(2\Delta)$
\cite{footnote2}.

Our goal is to select the time-dependent Rabi frequencies and laser phases  
\cite{kappa}
to accomplish the {\em ideal quantum transmission}
\begin{eqnarray}
\label{cond1}
&& \big(|g\rangle_1|\chi_g\rangle + |e\rangle_1 |\chi_e\rangle\big)
|g\rangle_2 \otimes |0\rangle_1 |0\rangle_2 |vac\rangle \nonumber\\
\rightarrow && |g\rangle_1 \big(|g\rangle_2 |\chi_g\rangle +
|e\rangle_2 |\chi_e\rangle\big) 
\otimes |0\rangle_1 |0\rangle_2
|vac\rangle,
\end{eqnarray}
where $|\chi_{g,e}\rangle$ are unnormalized states of other ``spectator''
atoms in the network. In (\ref{cond1}), $|0\rangle_i$ and $|vac\rangle$
represent the vacuum state of the cavity modes and the free electromagnetic
modes connecting the cavities.  Transmission will occur by photon
exchange via these modes.

In a quantum stochastic description employing the input--output
formalism the cavity mode operators obey the quantum Langevin
equations \cite{GardinerBook}:
\begin{equation}
\label{Lang}
\frac{d\hat a_i}{dt} = -i[\hat a_i,\hat H_{i}(t)] - \kappa \hat a_i
-\sqrt{2\kappa} \hat a_{\rm in}^{(i)}(t) \quad (i=1,2)
\end{equation}
The first term on the RHS of this equation gives the systematic
evolution due to the interaction with the atom, while the last two
terms correspond to photon transmission through the mirror with loss
rate $\kappa$, and (white) quantum noise of the vacuum field incident
on the cavity $i$, respectively.  The output of each cavity is given
by the equation \cite{GardinerBook}:
\begin{equation}
\hat a_{\rm out}^{(i)}(t)= \hat a_{\rm in}^{(i)}(t) + \sqrt{2\kappa}
\hat a_i(t)
\end{equation}
which expresses the outgoing field at the mirror as a sum of the
incident field plus the field radiated from the cavity. The output
field of the first cavity constitutes the input for the second cavity
with an appropriate time delay, i.e., $\hat a_{\rm in}^{(2)}(t)=\hat
a_{\rm out}^{(1)}(t-\tau)$, where $\tau$ is a constant related to
retardation in the propagation between the mirrors. The output field
of the second cavity is, therefore,
\begin{equation}
\hat a_{\rm out}^{(2)}(t)= \hat a_{\rm in}^{(1)}(t-\tau)
+ \sqrt{2\kappa} [\hat a_1(t-\tau) + \hat a_2(t)].
\end{equation}
Introducing this relation in Eqs.~(\ref{Lang}) we obtain
\begin{mathletters}
\label{Lang2}
\begin{eqnarray}
\frac{d\hat a_1}{dt} &=& -i[\hat a_1,\hat H_{1}(t)] - \kappa \hat a_1
-\sqrt{2\kappa} \hat a_{\rm in}^{(1)}(t)\\ \frac{d\hat a_2}{dt} &=&
-i[\hat a_2,\hat H_{2}(t)] - \kappa \hat a_2 -2\kappa \hat
a_1(t-\tau)\nonumber\\ && -\sqrt{2\kappa} \hat a_{\rm
in}^{(1)}(t-\tau)
\end{eqnarray}
\end{mathletters}
Note that the first equation is decoupled from the second one, i.e. we
consider here only a unidirectional coupling between the cavities (see
Fig.~1) \cite{footnote}. The present model is a particular example of
a cascaded quantum system and can be described within the formalism
developed by Gardiner and Carmichael \cite{Ga93,Ca93} .  We can eliminate
the time delay $\tau$ in these equations by defining ``time delayed''
operators for the first system (atom + cavity), e.g.~ $\tilde a(t)
\equiv \hat a(t-\tau)$, {\it etc.}; in a similar way we redefine the
Rabi frequency $\tilde \Omega_1(t)=\Omega_1(t-\tau)$, and phase
$\tilde \phi_1(t)=\phi_1(t-\tau)$. In the following we will assume
that we have performed these transformations, and for simplicity of
notation we will omit the tilde. This amounts to setting $\tau
\rightarrow 0$ in all these equations.  Equations (\ref{Lang2}) have
to be solved with the corresponding equations for the atomic operators
and with the condition that the field incident on the first cavity is
in the vacuum state, i.e.~ $\hat a_{\rm in}^{(1)}(t) |\Psi_0\rangle =
0$ $\forall t$.

In the present context, it is convenient to reformulate the above
problem in the language of quantum trajectories \cite{Ca93,Zo96}. Let
us consider a fictitious experiment where the output field of the
second cavity is continuously monitored by a photodetector (see
Fig.~1). The evolution of the quantum system under continuous
observation, conditional to observing a particular trajectory of
counts, can be described by a pure state wavefunction
$|\Psi_c(t)\rangle$ in the system Hilbert space (where the radiation
modes outside the cavity have been eliminated). During the time
intervals when no count is detected, this wavefunction evolves
according to a Schr\"odinger equation with non--hermitian effective
Hamiltonian
\begin{equation}
\label{Heff}
\hat H_{\rm eff}(t) = \hat H_1(t) + \hat H_2(t) - i \kappa \left( \hat
a_1^\dagger \hat a_1 + \hat a_2^\dagger \hat a_2 +2 \hat a_2^\dagger
\hat a_1 \right).
\end{equation}
The detection of a count at time $t_r$ is associated with a quantum
jump according to $|\Psi_c(t_r+dt)\rangle \propto \hat c
|\Psi_c(t_r)\rangle$, where $\hat c= \hat a_1 + \hat a_2$
\cite{Ga93,Zo96}.  The probability density for a jump (detector click)
to occur during the time interval from $t$ to $t+dt$ is $\langle
\Psi_c(t)| \hat c^\dagger\hat c |\Psi_c(t)\rangle dt$ \cite{Ga93,Zo96}.

We wish to design the laser pulses in both cavities in such a way that
ideal quantum transmission condition (\ref{cond1}) is satisfied.  A
necessary condition for the time evolution is that a quantum jump
(detector click, see Fig.~1) never occurs, i.e.~ $\hat c
|\Psi_c(t)\rangle =0$ $\forall t$, and thus the effective Hamiltonian
will become a hermitian operator. In other words, the system will
remain in a {\it dark} state of the cascaded quantum system.
Physically, this means that the wavepacket is not reflected from the
second cavity.  We expand the state of the system as
\begin{eqnarray}
\label{ansatz}
|\Psi_c(t)\rangle &=&
|\chi_g\rangle |gg\rangle |00\rangle \nonumber\\
&+& |\chi_e\rangle \Big[
\alpha_{1}(t) e^{-i\phi_1(t)} |eg\rangle |00\rangle
+ \alpha_{2}(t) e^{-i\phi_2(t)} |ge\rangle |00\rangle \nonumber\\
&& \hspace*{1cm}+ \beta_1(t) |gg\rangle |10\rangle
+ \beta_2(t) |gg\rangle |01\rangle\Big].
\end{eqnarray}
Ideal quantum transmission (\ref{cond1}) will occur for
\begin{equation}
\label{c}
\alpha_1(-\infty)=\alpha_2(+\infty)=1, \;
\phi_1(-\infty)=\phi_2(+\infty)=0.
\end{equation}
The first term on the RHS of (\ref{ansatz}) does not change under
the time evolution generated by $H_{\rm eff}$.
Defining symmetric and antisymmetric coefficients $\beta_{1,2}=
(\beta_s\mp \beta_a)/\sqrt{2}$, we find the following
{\em evolution equations}
\begin{mathletters}
\label{a}
\begin{eqnarray}
\label{a1}
\dot \alpha_{1}(t) &=& g_1(t) \beta_a(t)/\sqrt{2},\\
\label{a2}
\dot \alpha_{2}(t) &=& - g_2(t)\beta_a(t)/\sqrt{2},\\
\label{a3}
\dot \beta_a(t) &=& -  g_1(t) \alpha_{1}(t)/\sqrt{2} +
g_2(t) \alpha_2(t)/\sqrt{2}.
\end{eqnarray}
\end{mathletters}
where we have chosen the laser frequencies
$\omega_L+\dot \phi_{1,2}(t)$ so that $\delta=g^2/\Delta$ and
\begin{equation}
\dot \phi_{1,2}(t) = \delta\omega_i(t)
\end{equation}
in order to compensate the AC--stark shifts; thus Eq.~(\ref{a}) are
decoupled from the phases.
The {\em dark state condition} implies $\beta_s(t)=0$, and therefore
\begin{equation}
\label{b}
\dot \beta_{s}(t) =  g_1(t) \alpha_{1}(t)/\sqrt{2} +
g_2(t) \alpha_{2}(t)/\sqrt{2} + \kappa  \beta_{a}(t)
\equiv 0,
\end{equation}
as well as the normalization condition
\begin{equation}
\label{norm}
|\alpha_1(t)|^2 + |\alpha_2(t)|^2 + |\beta_a(t)|^2 = 1.
\end{equation}
We note that the coefficients $\alpha_{1,2}(t)$ and $\beta_s(t)$ are
real.

The mathematical problem is now to find pulse shapes $\Omega_{1,2}(t)
\propto g_{1,2}(t)$ such that the conditions (\ref{c},\ref{a},\ref{b})
are fulfilled.  In general this is a difficult problem, as imposing
conditions (\ref{c},\ref{b}) on the solutions of the differential
equations (\ref{a}) give functional relations for the pulse shape
whose solution are not obvious.  We shall construct a class of
solutions guided by the physical expectation that the time evolution
in the second cavity should reverse the time evolution in the first
one. Thus, we look for solutions satisfying the {\em symmetric pulse
condition}
\begin{equation}
g_2(t)=g_1(-t) \quad (\forall t).
\end{equation}
This implies $\alpha_{1}(t) = \alpha_2(-t)$, and
$\beta_a(t)=\beta_a(-t)$.  The latter relation leads to a symmetric
shape of the photon wavepacket propagating between the cavities.

Suppose that we specify a pulse shape $\Omega_1(t)\propto g_1(t)$ for
the second half of the pulse in the first cavity ($t\ge 0$)
\cite{footnote3}. We wish to determine the first half $\Omega_1(-t)
\propto g_1(-t)$ (for $t>0$), such that the conditions for ideal
transmission (\ref{cond1}) are satisfied. From (\ref{b},\ref{c}) we
have
\begin{equation}
\label{oo}
g_1(-t) = - \frac{\sqrt{2}\kappa \beta_a(t)
+g_1(t) \alpha_{1}(t)}{\alpha_2(t)}, \quad (t>0).
\end{equation}
Thus, the pulse shape is completely determined provided we know the
system evolution for $t\ge 0$. However, a difficulty arises when we
try to find this evolution, since it depends on the yet unknown
$g_2(t)= g_1(-t)$ for $t>0$ [see Eqs.(\ref{a})].  In order to
circumvent this problem, we use (\ref{b}) to eliminate this dependence
in Eqs.~(\ref{a1},\ref{a3}).  This gives
\begin{mathletters}
\label{backtothefuture}
\begin{eqnarray}
\dot \alpha_{1}(t) &=&
g_1(t)\beta_a(t)/\sqrt{2},\\
\dot \beta_a(t) &=& -\kappa \beta_a(t) - \sqrt{2}g_1(t) \alpha_{1}(t)
\end{eqnarray}
\end{mathletters}
for $t\ge 0$. These equations have to be integrated with the initial
conditions
\begin{mathletters}
\begin{eqnarray}
\alpha_{1}(0)&=&
\left[
\frac{2\kappa^2}{g_1(0)^2 + \kappa^2}
\right]^{\frac{1}{2}}\\
\beta_a(0) &=& \left[ 1- 2 \alpha_{1}(0)^2
\right]^{\frac{1}{2}}
\end{eqnarray}
\end{mathletters}
which follow immediately from $\alpha_1(0)=\alpha_2(0)$, and
(\ref{norm},\ref{b}) at $t=0$. Given the solution of
Eqs.~(\ref{backtothefuture}), we can determine $\alpha_2(t)$ from the
normalization (\ref{norm}). In this way, the problem is solved since
all the quantities appearing on the RHS of Eq.~(\ref{oo}) are known
for $t\ge 0$.  It is straightforward to find analytical expressions
for the pulse shapes, for example by specifying $\Omega_1(t)={\rm
const}$ for $t>0$. This is the pulse that will be considered in the
following discussion.

As an illustration, we have numerically integrated the full
time--dependent Schr\"odinger equation with the effective Hamiltonian
(\ref{Heff}). The results are displayed in Fig.~2(a).  We have used a
pulse shape calculated using the above procedure, with
$g_1(t)=2\delta\omega_1(t)=\kappa\equiv {\rm const}$ for $t>0$ [see
Fig.~2(b)]. As the figure shows, the quantum transmission is ideal.

In practice there will be several sources of imperfections. First,
there is the possibility of spontaneous emission from the excited
state during the Raman pulses. Its effects can be accounted for in the
effective Hamiltonian (\ref{Heff}) by the replacement $\Delta
\rightarrow \Delta+i\Gamma/2$, where $\Gamma$ is the decay rate from
level $|r\rangle$.  If we denote by $\tau$ ($\approx {\rm
max}(1/\kappa,1/g_{1,2})$) the effective pulse duration, the
probability for a spontaneous emission is of the order of $\Gamma
(\Omega_{1,2}^2+4g^2)/(8 \Delta^2) \, \tau \ll 1$. For $g_1 \approx
\kappa$ this probability scales like $1/\Delta$, so that the effects
of spontaneous emission are suppressed for sufficiently large
detunings. A second source of decoherence will be losses in the mirror
and during propagation. They can be taken into account by adding a
term $ -i \kappa' (\hat a_1^\dagger \hat a_1 + \hat a_2^\dagger \hat
a_2) $ in $H_{\rm eff}$ (\ref{Heff}), where $\kappa'$ is the
additional loss rate. Typically, we expect $\kappa' \ll
\kappa$. Nevertheless, one can overcome the effects of photon losses
by error correction \cite{error}. We have included these
imperfections in our numerical simulations. Fig.~3 shows the
probability of a faithful transmission $\cal F$ as a function of
$\kappa'/\kappa$ for different values of $\Gamma/\Delta$ for the same
parameters and pulse shapes as in Fig.~2.

In conclusion, we have proposed for the first time a protocol to
accomplish ideal quantum transmission between two nodes of a quantum
network.  Our scheme has been tailored to a potential network implementation
in which trapped atoms or ions constitute the nodes, and photon transmission
lines provide communication channels between them.  Extensions of the
present scheme will be presented elsewhere\cite{magic}, including error
correction and new quantum gates in cavity quantum electrodynamics.

We thank the members of the ITP program {\em Quantum Computers and
Quantum Coherence} for discussions. This work was supported in part by
the \"Osterreichischer Fonds zur F\"orderung der wissenschaftlichen
Forschung, by NSF PHY94-07194 and PHY-93-13668, by DARPA/ARO through
the QUIC program, and by the ONR.

% **********************************************************

% *************************************************************
\begin{figure}
%\begin{center}\
%{\psfig{file=Figure1.ps,width=8cm}}
%\end{center}
\caption{ Schematic representation of unidirectional quantum
transmission between two atoms in optical cavities connected by a
quantized transmission line (see text for explanation).  }
\end{figure}
% *************************************************************
% *************************************************************
\begin{figure}
%\begin{center}\
%{\psfig{file=Figure2.ps,width=5cm}}
%\end{center}
\caption{Populations $\alpha_{1,2}(t)^2$ and $\beta_a(t)^2$ for the
ideal transmission pulse $g_1(t)=g_2(-t)$ given in the inset,
specified by $g_1(t\ge 0)=2\delta\omega_1(t\ge 0)=\kappa={\rm const}$.  }
\end{figure}
% *************************************************************

% *************************************************************
\begin{figure}
%\begin{center}\
%{\psfig{file=Figure3.ps,width=5cm}}
%\end{center}
\caption{ Fidelity of a faithful transmission ${\cal F}$ including the
effects of a mirror losses and spontaneous emission as a function of
$\kappa'/\kappa$ for $\Gamma/\Delta= 0, 0.01, 0.05 $ (solid, dashed
and dot-dashed lines, respectively). The other parameters are as in
Fig.~2.  }
\end{figure}
% *************************************************************


\begin{references}

\bibitem{DiV95} See, for example, D. P. DiVincenzo, Science {\bf 270},
255 (1995).

\bibitem{Ci95} J. I. Cirac and P. Zoller, Phys. Rev. Lett. {\bf 74},
4091 (1995).

\bibitem{Pe95}
T. Pellizzari {\it et al}, Phys. Rev. Lett. {\bf 75 }, 3788 (1995).

\bibitem{Mo95}
C. Monroe {\it et al.}, Phys. Rev. Lett. {\bf 75}, 4714 (1995).

\bibitem{Tu95} Q. Turchette {\it et al.}, Phys. Rev. Lett. {\bf 75},
4710 (1995); for experimental CQED in the microwave regime see, for
example, M. Brune {\em et al.}., Phys. Rev. Lett. in press. %cat

\bibitem{Ma96}
K. Mattle {\em at al.}, Phys. Rev. Lett. {\bf 76}, 4656 (1996).

\bibitem{Be95}
C. H. Bennett, Phys.~Today {\bf 24} (October 1995);
A. K. Ekert, Phys. Rev. Lett. {\bf 67}, 661 (1991).

\bibitem{Be93}
C. H. Bennet {\it et al}, Phys. Rev. Lett. {\bf 70}, 1895 (1993).

\bibitem{Be96}
C. H. Bennet {\it et al.}, Phys. Rev. Lett. {\bf 76}, 722 (1996);
D. Deutsch  {\it et al.}, Phys. Rev. Lett. {\bf 77 }, 2818 (1996);
N. Gisin, Phys. Lett. A {\bf 210} 151 (1996),

\bibitem{magic}
J. I. Cirac {\it et al.}, in preparation.

\bibitem{La96}
C.K. Law and J.H. Eberly, Phys. Rev. Lett. {\bf 76}, 1055 (1996).

\bibitem{footnote2}
We ignore for the moment the small effects produced by spontaneous
emission during the Raman process. Its effects will be studied in
the context of Fig.~3.

\bibitem{kappa} One could also modulate the cavity transmission, but
this is technically more difficult.

\bibitem{GardinerBook}
C.W.~Gardiner,  {\em Quantum Noise} (Springer--Verlag, Berlin, 1991).

\bibitem{footnote}
In a perfect realization of the present scheme no light field will
be reflected from the second mirror, and therefore the assumption of
unidirectional propagation is not needed.

\bibitem{Ga93} C.W.~Gardiner, Phys.~Rev.~Lett.{\bf 70}, 2269 (1993).
\bibitem{Ca93} H.J.~Carmichael, Phys.~Rev.~Lett. {\bf 70}, 2273 (1993).

\bibitem{Zo96} For a review see
P. Zoller and C. W. Gardiner in {\em Quantum Fluctuations}, Les
Houches, ed. E. Giacobino {\it et al.} (Elsevier, NY, in press).

\bibitem{footnote3}
$ \Omega_1(t)$ has to be such that $\alpha_{1}(\infty)= 0$.
This is fulfilled if $ \Omega_1(\infty)>0$, which also guarantees
that the denominator in (\ref{oo}) does not vanish for $t>0$.

\bibitem{error} P.W.~Shor, Phys. Rev. A {\bf 52}, R2493 (1995);
A.M. Steane, Phys. Rev. Lett. {\bf 77}, 793 (1996); J. I. Cirac,
T. Pellizzari and P. Zoller, Science {\bf 273}, 1207 (1996); P. Shor,
{\em Fault--tolerant quantum computation}, quant--ph/9605011;
D. DiVincenzo and P.W. Shor, Phys. Rev. Lett. {\bf 77}, 3260 (1996).





\end{references}
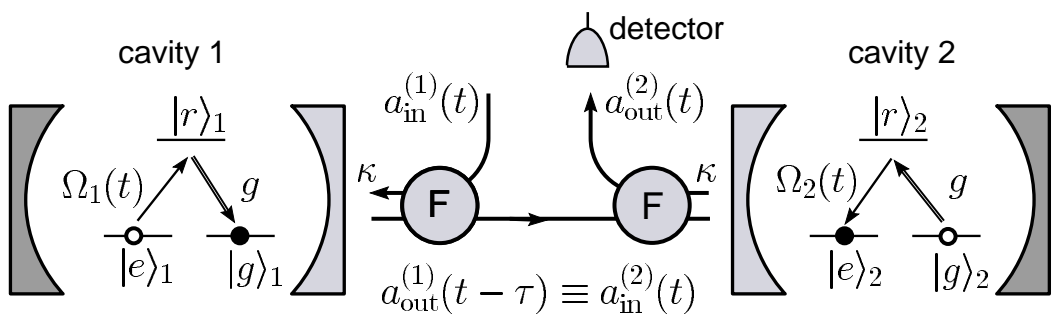
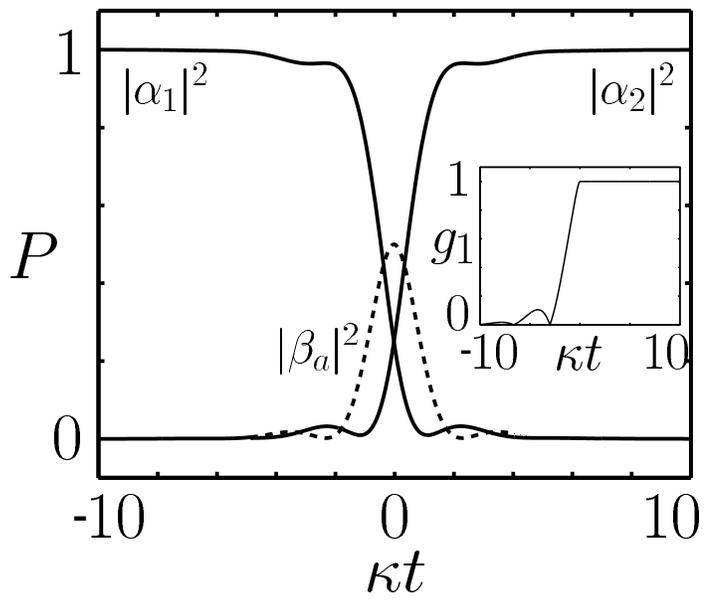
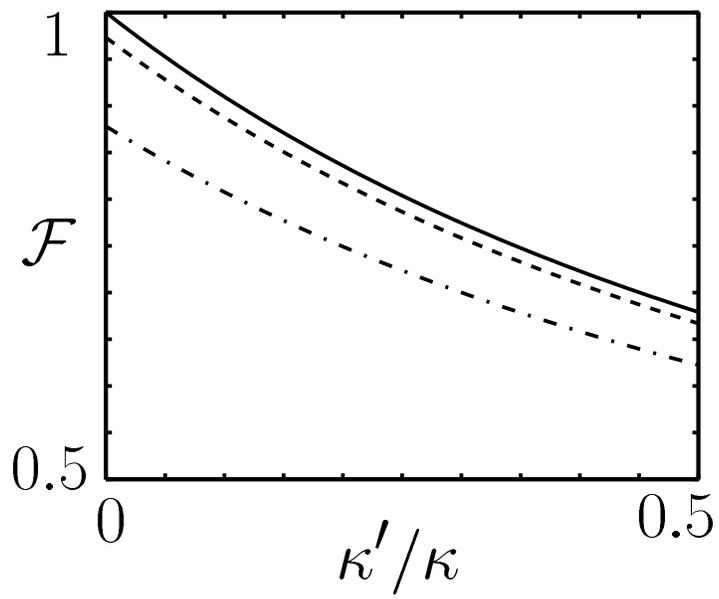
\end{document}